\documentclass[twocolumn]{article}

\usepackage{amsmath}
\usepackage{amsfonts}
\usepackage{amscd}
\usepackage{amssymb}
\usepackage{amsthm}
\usepackage{bm}
\usepackage{nicefrac}
\usepackage{wrapfig}

\usepackage[usenames,dvipsnames]{xcolor}

\usepackage{lineno}
\usepackage{graphicx}
\usepackage[labelfont=bf]{caption}
\usepackage[format=hang]{subcaption}

\usepackage[algoruled]{algorithm2e}
\usepackage{listings}
\usepackage{fancyvrb}
\fvset{fontsize=\normalsize}

\lstdefinestyle{mystyle}{
    commentstyle=\color{OliveGreen},
    keywordstyle=\color{BurntOrange},
    numberstyle=\tiny\color{black!60},
    stringstyle=\color{darkblue},
    basicstyle=\ttfamily,
    breakatwhitespace=false,
    breaklines=true,
    captionpos=b,
    keepspaces=true,
    numbers=left,
    numbersep=5pt,
    showspaces=false,
    showstringspaces=false,
    showtabs=false,
    tabsize=2
}
\usepackage{enumitem}

\usepackage{hyperref}
\usepackage[all]{hypcap}
\hypersetup{colorlinks,linktoc=all}
\hypersetup{citecolor=MidnightBlue}
\hypersetup{urlcolor=MidnightBlue}
\hypersetup{linkcolor=black}

\usepackage[nameinlink, capitalize]{cleveref}
\creflabelformat{equation}{#1#2#3}

\usepackage
[acronym,smallcaps,nowarn,section,nogroupskip,nonumberlist]{glossaries}
\glsdisablehyper{}
\setglossarystyle{long3col}
\makeglossaries

\usepackage{blindtext}
\usepackage{framed}

\usepackage{booktabs}
\usepackage{multirow}
\usepackage{longtable}
\usepackage{etoolbox,siunitx}
\robustify\bfseries
\DeclareRobustCommand{\parhead}[1]{\textbf{#1}~}

\usepackage{appendix}
\Crefname{appsec}{appendix}{appendices}
\AtBeginEnvironment{appendices}{\crefalias{section}{appendix}}

\usepackage{algorithmic}

\usepackage{tablefootnote}
\usepackage{breqn}

\usepackage{stmaryrd}

\newcommand*{\E}[2][]{\mathbb{E}\ifx\\\left[#1\right]\\\else_{#1}\fi \left[#2\right]}

\usepackage{stmaryrd}
\newcommand{\abs}[1]{\left| #1 \right|}
\newcommand{\parens}[1]{\left( #1 \right)}

\newcommand{\bbrackets}[1]{\llbracket #1 \rrbracket}
\newcommand{\R}{\mathbb{R}}

\newtheorem{bound}{Bound}
\newtheorem{lemma}{Lemma}

\newtheorem{remark}{Remark}

\newtheorem*{example*}{Example}

\DeclareMathOperator*{\argmin}{arg\,min}
\usepackage{dsfont}

\newif\ifmtpro
\newif\ifsimple
\newif\ifhipster

\mtprotrue

\ifmtpro
\usepackage[T1]{fontenc}
\usepackage[full]{textcomp}
\usepackage{newtxtext}
\usepackage{cabin} 
\usepackage[english]{babel}

\usepackage{bm}
\fi

\ifsimple
\usepackage[T1]{fontenc}
\usepackage{ccfonts}
\usepackage{amssymb}
\usepackage[bb=mt,cal=mt]{mathalpha}
\fi

\ifhipster
\usepackage{ccfonts}
\usepackage[T1]{fontenc}
\usepackage{textcomp}
\usepackage[final]{microtype}
\usepackage[english]{babel}

\usepackage{amssymb}
\usepackage{mathbbol}
\usepackage{upgreek}

\usepackage{everysel}
\EverySelectfont{
  \fontdimen2\font=0.4em
  \fontdimen3\font=0.2em
  \fontdimen4\font=0.1em
  \fontdimen7\font=0.1em
  \hyphenchar\font=`\-
}
\usepackage{bm}

\fi

\usepackage{amsfonts}
\usepackage{eucal}
\usepackage{amsmath}
\usepackage{amsthm}
\usepackage{thmtools}
\usepackage{mathtools}

\usepackage[parfill]{parskip}
\usepackage{afterpage}
\usepackage{setspace} 

\RequirePackage[bf,tiny]{titlesec}

\usepackage[
paper  = letterpaper,
left   = 1.25in,
right  = 1.25in,
top    = 1.0in,
bottom = 1.0in,
]{geometry}

\usepackage[%
minnames=1,maxnames=99,maxcitenames=2,	
style=alphabetic,
doi=false,	
url=false,	
natbib,	
backend=bibtex,	
sorting=nyt	
]{biblatex}%

\DeclareFieldFormat{sentencecase}{\MakeSentenceCase*{#1}}	

\renewbibmacro*{title}{%
  \ifthenelse{\iffieldundef{title}\AND\iffieldundef{subtitle}}	
    {}	
    {\ifthenelse{\ifentrytype{article}\OR\ifentrytype{inbook}%
      \OR\ifentrytype{incollection}\OR\ifentrytype{inproceedings}%
      \OR\ifentrytype{inreference}}	
      {\printtext[title]{%
        \printfield[sentencecase]{title}%
        \setunit{\subtitlepunct}%
        \printfield[sentencecase]{subtitle}}}%
      {\printtext[title]{%
        \printfield[titlecase]{title}%
        \setunit{\subtitlepunct}%
        \printfield[titlecase]{subtitle}}}%
     \newunit}%
  \printfield{titleaddon}}

\AtEveryBibitem{%
\ifentrytype{article}{	
    \clearfield{url}%
    \clearfield{urldate}%
    \clearfield{eprint}	
    \clearfield{eid}	
}{}	
\ifentrytype{book}{	
    \clearfield{url}%
    \clearfield{urldate}%
}{}	
\ifentrytype{collection}{	
    \clearfield{url}%
    \clearfield{urldate}%
}{}	
\ifentrytype{incollection}{	
    \clearfield{url}%
    \clearfield{urldate}%
}{}	
}	

\AtEveryBibitem{	
    \clearfield{pages}	
    \clearfield{review}%
    \clearfield{series}
    \clearfield{volume}	
    \clearfield{month}	
    \clearfield{eprint}	
    \clearfield{isbn}	
    \clearfield{issn}	
    \clearlist{location}	
    \clearfield{series}	
    \clearlist{publisher}	
    \clearname{editor}	
}{}	

\usepackage{lipsum}
\usepackage{xr}

\usepackage[affil-it]{authblk}
\title{On the Misspecification of Linear Assumptions in Synthetic Control}
\author{Achille Nazaret}
\author{Claudia Shi}
 \author{David M. Blei}
\affil{Columbia University}
 \date{}
\title{On the Misspecification of Linear Assumptions in Synthetic Control}

\begin{document}
\maketitle

\begin{abstract}
The synthetic control (SC) method is a popular approach for estimating treatment effects from observational panel data.  It rests on a crucial assumption that we can write the treated unit as a linear combination of the untreated units.  This linearity assumption, however, can be unlikely to hold in practice and, when violated, the resulting SC estimates are incorrect. In this paper we examine two questions: (1) How large can the misspecification error be?  (2) How can we limit it?  First, we provide theoretical bounds to quantify the misspecification error.  The bounds are comforting: small misspecifications induce small errors.  With these bounds in hand, we then develop new SC estimators that are specially designed to minimize misspecification error.  The estimators are based on additional data about each unit, which is used to produce the SC weights.  (For example, if the units are countries then the additional data might be demographic information about each.) We study our estimators on synthetic data; we find they produce more accurate causal estimates than standard synthetic controls.  We then re-analyze the California tobacco-program data of the original SC paper, now including additional data from the US census about per-state demographics.  Our estimators show that the observations in the pre-treatment period lie within the bounds of misspecification error, and that the observations post-treatment lie outside of those bounds.  This is evidence that our SC methods have uncovered a true effect.

\end{abstract}

\section{Introduction}\label{sec:intro}

The synthetic control (SC) method is a popular approach for analyzing observational panel data to estimate causal effects \citep{abadie2021using}.  SC has been widely used in science \citep{pieters2016effect}  and social science \citep{heersink2017disasters}, as well as for evaluating public policies \citep{donohue2019right,pinotti2015economic, allegretto2017credible}.\looseness=-1

The typical SC setup involves measurements of an outcome variable over time. One unit, called the \textit{target}, received an intervention at a certain time. The other units, called \textit{donors}, never received an intervention.  The goal of SC is to estimate the target's counterfactual outcomes. What would have happened had it not received the intervention?

\begin{figure}
    \centering
    \includegraphics[width=\linewidth]{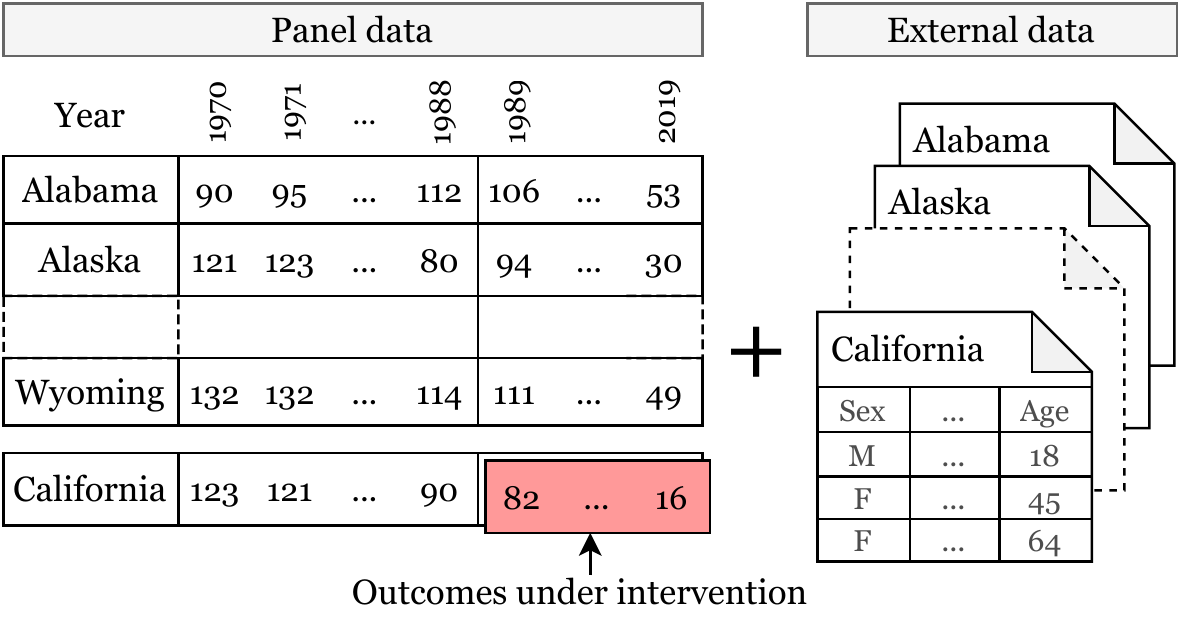}
    \caption{Representation of the data used by our estimators. In addition to panel data of annual per-capita cigarette sales (in packs), our estimators leverage external data containing demographic information about each state.}
    \label{fig:data}
\end{figure}

Example: The panel data in \cref{fig:data} (left) contains cigarette sales across states and time. In 1988 California implemented a program that increased the tobacco tax by 25 cents. After 1988, how much would Californians have smoked had the program not been implemented? Here, \mbox{California} is the target; the other states are the donors.

The idea behind SC is to approximate the target's control
outcomes---the smoking rate in California without its policy---with a
weighted combination of the donor's control outcomes. In the example,
SC uses data from the pre-policy periods to fit California's
pre-policy smoking rates as a weighted combination of the other
states' smoking rates. It then uses its fitted weights to estimate the
smoking rate in California after 1988, had the policy not been
introduced. These estimates, along with California's post-policy
rates, help assess the causal effect of the policy.

What justifies this procedure?  In its original formulation, \citet{abadie2010synthetic} shows that SC is justified if the control outcomes follow a linear factor model, where a per-period factor linearly combines with a per-unit factor.  Following this work, \citet{shi2022assumptions} shows that the linear factor model itself can be justified through assumptions about the individuals within each unit (e.g., people within each state) and invariances around the causal structure of the individual-level outcomes (e.g., whether they smoke).  But whether at the aggregate or individual level, these assumptions point to the same requirement: that the target need to be expressed as a linear combination of the donors.

What if this requirement is not satisfied? What if California is not a linear combination of the other states? This paper studies the practical situation where the synthetic control is \textit{misspecified}. We study how to quantify this misspecification error and how to minimize it.

In detail, we derive two bounds on the SC error, the M bound and the
James Bound. Both bounds build on the causal framework of
\citet{shi2022assumptions}. It assumes a set of \textit{invariant
  causes}, variables that govern the individual-level outcomes in the
same way across units, and where the difference between the units'
outcomes involves different distributions of those causes. For
example, whether someone smokes might be caused by their age and
education level, and the difference between California's and Nevada's
smoking rates lies in their different population distributions of
those demographic variables. Our theory shows how the similarity
between the true target distribution of the causes and its synthetic
distribution, induced by the SC weights, helps bound the error of the
corresponding SC estimates.

We then consider a situation where we additionally observe external
data about the invariant causes, such as demographic information about
each state. We show how to use such data to estimate the
misspecification interval for a fixed set of SC weights, and we
develop two new algorithms for estimating SC weights that explicitly
minimize the width of this interval. (One algorithm assumes we observe
all invariant causes; the other does not make that assumption, but
provides wider misspecification intervals.)

Thus this paper provides a new form of SC analysis, one where we analyze panel data and demographic data together to estimate the target counterfactual and assess its robustness to misspecification. \Cref{fig:california_jamesbound} illustrates this analysis on the California tobacco data, now also using additional data from the U.S. census about per-state demographics. Our estimators show that the observed outcome in the pre-policy period lies within the bounds of misspecification error, and that the observed outcome post-policy lies outside of those bounds. These results suggest that California's 1988 anti-tobacco program had a true effect, despite possible misspecification of the synthetic control.


\parhead{Related Work.} This paper contributes to the literature on synthetic controls~\citep{abadie2003economic, abadie2010synthetic}. The M-bound and James-bound estimators of \Cref{sec:m-bound,sec:jamesbound} contribute to research on novel SC estimators \citep{abadie2011bias, abadie2015comparative,  doudchenko2016balancing, xu2017generalized, amjad2018robust, amjad2019mrsc, li2020statistical, athey2021matrix, imbens2021controlling}. Notably, several estimators penalize the optimization objective \citep{abadie2021penalized} or adjust the weights \citep{kellogg2021combining} to select donors with outcomes similar to the target. Including covariates in the estimator is recommended in \citet{abadie2003economic} but is not mathematically justified. In contrast, the M and James bounds justify why selecting similar donors is important for estimation.

This paper also contributes to the literature on SC methods that are robust to misspecifications.
Many existing works make additional assumptions; for example, outcomes follow linear factor models \citep{ben2021augmented},
latent factors are perfectly matched \citep{amjad2018robust,  powell2018imperfect, ferman2021synthetic},
treatments are assigned randomly  \citep{abadie2015comparative},
or many time points are observed \citep{chernozhukov2021exact}.
In contrast, we leverage external data to quantify the errors.


\section{Synthetic Controls and Misspecification}\label{sec:background}

We review the assumptions behind synthetic controls, the fine-grained model of \citet{shi2022assumptions}, and formulate the problem of how to characterize the misspecification error induced by violation of the linearity assumptions.

\subsection{Panel Data and a Causal Question}
Consider a panel dataset containing outcome measurements $y_{jt}$ for units $j\in \bbrackets{0, J}$ over time periods $t\in\bbrackets{0, T}$. Unit $j=0$ is the \textit{target}. It has received an intervention at $T_0$ that may have affected its outcomes $y_{0t}$ for $t \geq T_0$. The remaining units are \textit{donors}. They did not receive an intervention.

For each unit and time, define a pair of potential outcomes $(Y_{jt}, \widetilde{Y}_{jt})$, where $\widetilde{Y}_{jt}$ is the potential outcome of unit $j$ at time $t$ in the world where $j$ received intervention at $T_0$, while $Y_{jt}$ is the potential outcome in a world with no intervention. 
For $j=0$ and $t \geq T_0$,  $y_{jt} =\widetilde{Y}_{jt}$, otherwise, $y_{jt} = Y_{jt}$.

Our causal question is, what would the target counterfactual be, had the intervention not occurred? We would like to estimate $Y_{0t}$ for $t \geq T_0$.


\subsection{Synthetic Controls and their Assumptions}

Synthetic control methods estimate the counterfactual outcomes of the target $Y_{0t}, t \geq T_0$ with a weighted combination of the outcomes of the donors: $Y_{0t} = \sum_j w_j y_{jt}$. The SC weights are fitted from the pre-intervention outcomes,
\begin{align}
w = \argmin_{w \in \Delta^J} \sum^{T_0-1}_{t=0} \Bigl(y_{0t} - \sum_{j}{w}_j  y_{jt} \Bigr)^2.  \label{eq:estimator}
\end{align}

The validity of SC relies on two conditions: (1) During the pre-intervention period, the target's outcomes can be written as a weighted combination of the control units' outcomes. (2) The weighted combination from the pre-intervention periods generalizes to the post-intervention periods.

As a first step to obtain these two conditions, \citet{abadie2010synthetic} and most other works assume that the outcomes under no intervention are generated by a linear factor model \citep{bai2009panel, abadie2010synthetic, ben2021augmented,  ferman2021synthetic}. We call it assumption A1.

\begin{description}[leftmargin=0cm]
\item[A1. (Linear Factor Model)]
Under no intervention, the outcomes are generated from a linear factor model,
\begin{equation}
    Y_{jt} = \mu_j ^\top \lambda_t + \epsilon_{jt}, \label{eq:linear_model}
\end{equation}
where $\mu_j$ is a unit-specific latent factor, $\lambda_t$ is a time-dependent factor, and $\epsilon_{jt}$ is independent random noise. 
\end{description}

Then, \citet{abadie2010synthetic} assumes that the target's outcomes can be written as a convex combination of the donors' outcomes. This implies that the target's latent factor is a convex combination of the donors' latent factors, we it call A2.\looseness=-1
\begin{description}[leftmargin=0cm]
\item[A2. (Convex Combination)]
The target unit's latent factor is a convex combination of the donors' latent factors,
\begin{equation*}\textstyle
    \exists w \in \Delta^J,  \hspace{2em} \mu_{0} = \sum_j w_j \mu_{j},\label{eq:convex}
    \end{equation*}
    where $\Delta^J$ is the simplex over $J$ coordinates.
\end{description}
\begin{remark}\label{rm:linear}
To be precise, \citet{abadie2010synthetic} assumes that $A:=\sum_{t < T_0} \lambda_t^\top \lambda_t$ is nonsingular. Consequently, once they assume $Y_{0t} = \sum_j w_j Y_{jt}$ for $t<T_0$, then the invertability of $A$ implies $\mu_0 = \sum_j w_j \mu_j$ (intuitively, we ``invert'' the factor model). The target is a convex combination of donors.
\end{remark}

With assumptions A1 and A2 in hand, estimator (\ref{eq:estimator}) will identify the weights of A2. These weights will then estimate the untreated potential outcomes using the factor model A1.
With assumptions A1 and A2, synthetic control is possible.

\subsection{A Fine-grained Model for SC}
\begin{figure}
    \centering
    \includegraphics[width=0.95\linewidth]{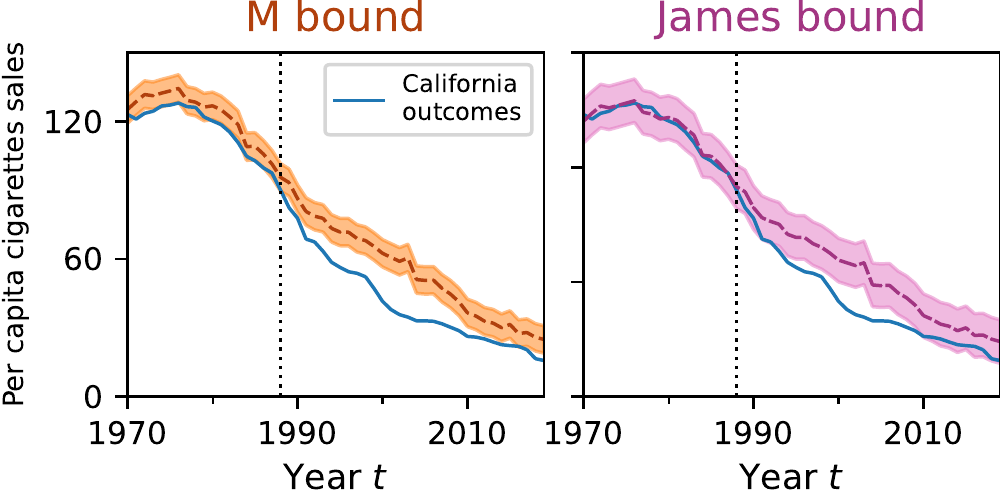}
    \caption{
Visualization of the observed California outcomes (solid lines), the SC estimates (dotted lines), and the misspecification intervals (shaded area), as calculated by the M-bound estimator (left) or the James-bound estimator (right).
 California's outcomes lie within the error bounds prior to the intervention but escape the error bounds after the intervention. This suggests that the tobacco program had a causal effect despite possible misspecification of SC.
}    \label{fig:california_jamesbound}
\end{figure}
When does the linear factor model assumption A1 hold in practice? \citet{shi2022assumptions} explores a justification. 

First, the authors notice that SC often considers large units composed of multiple individuals (states, countries) and aggregated outcomes that are averages of individual-level outcomes (per-capita cigarette consumption). Therefore, they propose a ``fine-grained" model of synthetic controls, which introduces individual-level variables. The variable $Y_{ijt}$ denotes individual $i$'s outcome in unit $j$ at time $t$ (their cigarette consumption at time $t$). The outcome of each unit $Y_{jt}$ is the average of the individual outcomes in the unit. 

Second, \citet{shi2022assumptions} posits the idea of \emph{invariant causes}, which we denote as $x_{ijt}$. The invariant causes $x_{ijt}$ are individual-level variables that follow two invariance assumptions.
(1) When conditioned on the invariant causes, the individual-level outcomes do not depend on which unit the individual is from. This way, $x \mapsto \E{Y_{ijt} | x_{ijt}=x}$ is the same function for all $i$ and $j$; we write it $\E[t]{Y | x}$.
(2) The distribution of the invariant causes in each unit can change from unit to unit but remain the same across time. The distribution of invariant causes over the individuals in unit $j$ is denoted $x \mapsto p_j(x)$, with no dependence in $t$. 

Finally, with these individual-level variables, \citet{shi2022assumptions} shows that the unit-level outcomes are
\begin{align}
\E{Y_{jt}} = \int_x \E[t]{Y \mid x} p_{j}(x)~ \text{d}x \label{eq:shi}.
\end{align}
Further, if the distributions of invariant causes $p_j$ are \emph{discrete} and \emph{finite} then \cref{eq:shi} becomes a finite sum, \begin{align}
   \E{Y_{jt}} = \sum_{x_k} \underbrace{\E[t]{Y \mid x_k}}_{(\lambda_t)_k}\underbrace{p_j(x_k)}_{(\mu_j)_k} = \lambda_t ^\top \mu_j.
   \label{eq:shi-2}
\end{align}
Thus \Cref{eq:shi-2} justifies the linear factor model A1 and provides  context about what the latent factors might represent. 

Regarding A2, \citet{shi2022assumptions} still assumes it; it rewrites,
\begin{equation*}\textstyle
    \exists w \in \Delta^J,  \hspace{2em} p_{0} = \sum_j w_j p_{j},\label{eq:convex_p}
    \end{equation*}
where the equality is in the space of probability distributions.

\subsection{Misspecification in SC}\label{sec:problem}

\citet{shi2022assumptions} explains where the linear factor model A1 can come from. However, even with the fine-grained approach, the factor model arises from the strong assumption of \textit{discrete} and \textit{finite} causes, which might not hold in practice. In other words, the factor model A1 is not guaranteed.

The convex combination A2 is also a key assumption that is unlikely to hold in practice. With a limited number of donors, the $p_j$ (or $\mu_j$) can be linearly independent. And with continuous invariant causes, the $p_j$ are densities, making it impossible to match $p_0$ even with infinitely many donors.

In this work, we relax A1 and A2. To relax A1, we use the fine-grained model of \citet{shi2022assumptions} but do not assume that causes are \textit{discrete} and \textit{finite}. 
To relax A2, is the focus of the paper.
When A2 is violated, the target is not a convex combination of the donors. The synthetic control is \textit{misspecified}, which leads to errors in the estimation of causal effects. Formally, we define the misspecification error as the absolute difference between the expected counterfactual outcome and the synthetic outcome (after intervention): \begin{center}
$
 \left| \E{Y_{0t}} -  \sum_{j=1}^J w_j \E{Y_{jt}} \right|$  for $t \geq T_0$.
\end{center}



We turn to the problem of how to bound and minimize this error. We will show how to leverage external data about the invariant causes -- such as demographic data from the census -- to estimate a \textit{misspecification interval} for the synthetic outcomes. We then derive new ways to fit the SC weights that minimize the width of that interval. The result is a new type of estimate of the SC counterfactual, and an assessment of its sensitivity to the linearity misspecification of A2.

\section{The M Bound and its Estimator}\label{sec:m-bound}
In this section, we derive an exact bound that quantifies errors induced by violation of A2.
Using this bound, we develop an estimator minimizing A2 misspecification errors.

\subsection{The M Bound}

Let $\hat{p}_0$ be the synthetic distribution, defined as $\hat p_0(w) = \sum_j w_j p_j$. We examine the difference between the  distribution of the target unit $p_0$ and the synthetic distribution $\hat p_0$. 
If $p_0 \neq \hat p_0$ but $\hat p_0$ remains ``close'' to $p_0$, we expect the synthetic control estimate to remain approximately correct,
\[ \E{Y_{0t}} \approx \sum_{j=1}^J w_j \E{Y_{jt}}.
\]
We formalize this intuition by bounding the errors resulting from the misspecification of A2. 



\begin{bound}[M bound]
\label{theorem:bound1}
For any $t$, assume that $x \mapsto \E[t]{Y | x}$ is $\ell$-Lipschitz\footnote{See \cref{appsec:technical} for details about Lipschitz functions.}. Then for any weights in the simplex $w$, we have the Misspecification error bound (M-bound):
\begin{align}
\abs{
    \E{Y_{0t}} - \sum_{j=1}^J w_j \E{Y_{jt}}
} \leq  \ell \cdot W_1\parens{p_{0}, \hat p_0}, \label{eqn:bound1}
\end{align}
where $\hat p_0 = \sum_j w_j p_{j}$ and $W_1$ is a $\ell_1$-Wasserstein \mbox{distance}.
\end{bound}
The proof is in \cref{appsec:technical}.

The Wasserstein distance $W_1$ is a distance between probability distributions \citep{villani2009optimal}. It quantifies the differences between the true population distribution $p_0$ and the synthetic population distribution $\hat p_0$.

For any set of weights $w$, the M bound (\Cref{theorem:bound1}) circumscribes the error of the SC estimate by a function of the weights, the population distributions of each unit (the $p_j$), and the sensitivity of the outcome variables to the variation of the causes (the Lipschitz constant $\ell$).

If $p_0 = \hat p_0$, then the Wasserstein distance $W_1\bigl(p_0, \hat p_0)$ between the true and the synthetic distribution is zero.
The M bound recovers that the SC estimate is correct. 
When $p_0 \neq \hat p_0$, the M bound shows that the estimation error is proportional to the distance $W_1\bigl(p_0, \hat p_0)$. 

The intuition behind \cref{eqn:bound1} is that when a misspecification occurs, a portion of the population $p_0$ is approximated with an incorrect portion of the synthetic population $\hat p_0$. It is unpredictable how these populations will behave. In the worst case, their outcomes can differ by at most the distance between them (captured by $W_1$) and the maximum possible variation of the conditional outcome (captured by $\ell$). Hence, the M bound proves (theoretically) that a small misspecification induces a small estimation error. 

\subsection{The M-bound Estimator}

We established the M bound, which quantifies the misspecification error for any set of weights $w$. To find weights with minimal misspecification error, we develop the M-bound estimator. See \cref{alg:mbound}.

The M-bound estimator takes population distribution data $p_j$ for each unit as input and returns a set of weights that minimizes the M bound.
To obtain the weights, it uses projected gradient descent with the following objective,  \begin{equation*}
    (w_1, ..., w_J) \mapsto W_1\Bigl(p_0, \sum_j w_j p_j\Bigr).
\end{equation*}



\begin{algorithm}[t]
   \caption{Minimization of the M bound}
   \label{alg:mbound}
\begin{algorithmic}
   \STATE {\bfseries Input:} Distributions $p_0, ..., p_J$; learning rate $\alpha$; number of epochs $E$.
   \STATE {\bfseries Output:} $(w_j)$ minimizing the M bound. 
   \STATE $(w_1, ..., w_J) \leftarrow \bigl(\frac1J, ..., \frac1J\bigr) $
   \FOR{$e=1$ {\bfseries to} $E$}
       \STATE $\hat p_0 \leftarrow \sum w_j p_j$
       \STATE $\text{grad} \leftarrow \nabla_{w} W_1\bigl(p_0, \hat p_0 \bigr) $
       \STATE $w \leftarrow w - \alpha \cdot \text{grad}$
       \STATE $w \leftarrow \text{project\_simplex}(w) $
   \ENDFOR
  \STATE \textbf{return} $w$
\end{algorithmic}
\end{algorithm}

Notice it computes the SC weights using the population distribution of each unit. It does not use the outcomes data.

After obtaining a set of weights from \cref{alg:mbound}, we can use \cref{eqn:bound1} with an estimated constant $\ell$ to create a misspecification interval around the synthetic control estimate,
\begin{equation}
\textstyle
 \E{Y_{0t}} \in  \left[\hat y_{0t} - M, \hat y_{0t}  + M  \right] \hspace{2em} \forall t,
\label{eq:m-interval}
\end{equation}
where $\hat y_{0t} := \sum_{j=1}^J w_j y_{jt}, ~M := \ell \cdot W_1\parens{p_{0}, \hat p_0}$. Thus, the M bound, with its associated estimator and misspecification interval, can be used to discover causal effects.

In \cref{subsec:case-study}, we revisit the California tobacco example.
We use demographic data of each US state to form the invariant causes distributions $p_j$ and fit the M-bound estimator with these $p_j$. Like standard SC, the weights returned by the estimator are used to form the synthetic outcomes. In addition, the M bound provides misspecification intervals accounting for the A2 misspecification error.
\cref{fig:california_jamesbound} illustrates the synthetic control estimate with its misspecification interval generated by the M-bound estimator.
We see that California's observed outcomes lie within the interval before intervention and escape it after the intervention. This suggests that a causal effect is present, even in case of misspecification.\looseness=-1

\section{The James Bound and its Estimator}\label{sec:jamesbound}
In \cref{sec:m-bound}, we derived a theoretical bound on misspecification error and showed how to use the M-bound estimator to detect a causal effect.
In theory, the true outcome is guaranteed to lie within the M bound.
In practice, the misspecification interval produced by the M bound is only valid if we observe the distribution of all invariant causes $p_j$.
Observing all invariant causes is a strong assumption that may not hold. 

Here, we consider the setting where the invariant causes are only partially observed.
We first derive a new error bound, the James bound, that accounts for misspecification on both the observed and unobserved causes.
The James bound leverages the pre-intervention outcome data to estimate the influence of the unobserved causes on the outcome variable. 
To find the weights that minimize the James bound, we develop the James-bound estimator.
Finally, we discuss when it is appropriate to use the M bound versus the James bound.\looseness=-1
\subsection{The James Bound}

So far, we have used $x$ to denote all the invariant causes. With a redefinition of notation, we now refer to the \emph{observed causes} as $x$, and the \emph{unobserved causes} as $z$. Such that \cref{eq:shi} becomes $\E{Y_{jt}} = \int_{(x,z)}p_j(x,z) \E[t]{Y|x,z}\text{d}x\text{d}z$. 

In general, we cannot bound the effect of unobserved variables without further assumptions.
Here, we assume that the unobserved causes and observed causes are independent and that their respective effect on the outcome can be decomposed into two distinct terms, this is A3. We note that A1, which we relaxed, was more restrictive than A3.


\begin{description}[leftmargin=0cm]
\item[A3. Independence of Observed and Unobserved Causes.] For each unit $j$, the variable $x$ and $z$ are independent,
\begin{equation*}
    p_j(x,z) = p_j(x)p_j(z),
\end{equation*}
and for each time $t$, there exist functions $g$ and $h$ such that:
\begin{equation*}
    \E[t]{Y | x, z} = g_t(x) + h_t(z).
\end{equation*}
\end{description}
We note that the distributions of the observed causes $x\mapsto p_j(x)$ and the unobserved causes $z\mapsto p_j(z)$ remain arbitrary, and so are $g_t$ and $h_t$. 

With A3, we have ``just another misspecification error'' bound, the James bound. 
\begin{bound}[James bound]
\label{theorem:bound2}
For $t \geq T_0$, assume that $x \mapsto \E[t]{Y | x}$ is $\ell$-Lipschitz. Then for any weights $w \in \Delta^J$,
\begingroup
\medmuskip=3mu
\thinmuskip=2mu
\thickmuskip=2mu
\begin{equation}
\textstyle
\abs{
    \E{Y_{0t}} - \sum_{j=1}^J w_j \E{Y_{jt}}
} \leq 
 \ell \cdot W_1(p_0(x), \hat p_0(x)) \label{eqn:bound2-1}
\end{equation}
\endgroup
\begin{equation}
\textstyle
~~~~~~~~~~~~~~~~~~~~~ +\max_{u < T_0} \abs{\E{Y_{0u}} - \sum_{j=1}^J w_j \E{Y_{ju}}} \label{eqn:bound2-2}
\end{equation}
\begingroup
\medmuskip=-1mu
\thinmuskip=-2mu
\thickmuskip=-1mu
\begin{equation}
\hspace*{-0.1cm}+\hspace*{-0.1cm}\inf_{\hspace*{-0.1cm}\alpha \in \Delta^{ T_0}} \abs{\int_{z}\hspace*{-0.08cm} \textstyle  \Bigl(\hspace*{-0.05cm}p_0(z) - \hat p_0(z)\hspace*{-0.07cm}\Bigr) \hspace*{-0.1cm}\Bigl(\E[t]{Y | z} - \hspace*{-0.12cm} \sum\limits_{u<T_0} \hspace*{-0.05cm} \alpha_u \E[u]{Y | z} \Bigr) \textnormal{d}z}. \label{eqn:bound2-3}
\end{equation}
\endgroup
\end{bound}
The proof is in \cref{appsec:technical}.


The first term (\ref{eqn:bound2-1}) mirrors the M bound. It quantifies the similarity between the target and synthetic unit's distributions of observed causes, $x\mapsto p_0(x)$ and $x\mapsto \hat p_0(x)$. 

The second term (\ref{eqn:bound2-2}) measures the goodness of fit of the pre-intervention outcomes. 
It indirectly estimates the similarity between the target and the synthetic unit's distributions of unobserved causes, $z\mapsto p_j(z)$ and $z\mapsto \hat p_j(z)$.


The last term (\ref{eqn:bound2-3}) contains the remaining error terms. We cannot compute this term directly because it contains unobserved quantities. In \cref{appsec:technical}, we argue that this term is small, and we may ignore it in practice.  We also present two explicit models in which the term is null and show that it is null in the standard SC factor model. 

\subsection{The James-bound Estimator}
Building on the James bound (\cref{theorem:bound2}), we derive the James-bound estimator.
The estimator identifies the weights that minimize the following objective,
\begin{equation}\textstyle
    w \mapsto \max\limits_{t < T_0} \Bigl|y_{0t} - \sum\limits_{j=1}^J w_j  y_{jt}\Bigr|  + \lambda \cdot W_1\Bigl(p_0, \sum_j w_j p_j\Bigr), 
    \label{eq:james-estimator}
\end{equation}
where $\lambda$ is a hyperparameter. We update \cref{alg:mbound} into a minimization algorithm for this objective function, it is reported in \cref{appsec:technical}.

If hyperparameter $\lambda$ is set to $\ell$, and if term (\ref{eqn:bound2-3}) is effectively negligible, 
then \cref{eq:james-estimator} is precisely the James bound. Otherwise, it can be viewed as the pre-intervention errors (first term), regularized by the Wasserstein distance over external data (second term). 
With this perspective, the James-bound estimator finds weights that minimize pre-intervention errors while favoring donors that are similar to the target.\looseness=-1

\subsection{Choosing between M and James Bound}\label{sec:choosing}
We introduced two bounds, along with associated estimators and misspecification intervals. M bounds are tighter but require data about all the invariant causes. James bounds can be wider but require fewer data. 

As a practical guide, we recommend using the James-bound estimator first. It is indeed more prudent to assume that some invariant causes might be unobserved. If the post-intervention target outcomes fall outside the misspecification interval, we have discovered a causal effect robust to A2 misspecification (see \cref{fig:california_jamesbound}).  

If the post-intervention misspecification interval is too wide to detect a causal effect, then it could be that there is no causal effect. But it could also be that there is too much misspecification to use SC or that the James bound is too loose. We cannot conclude in favor of a causal effect in the first two cases. To check if the James bound is too loose and find a tighter bound, we can use the M bound.

The M bound is guaranteed valid if all invariant causes are observed. Since the M estimator does not use outcome data, the target's pre-intervention outcomes can be used as a validation set. If the observed pre-intervention outcomes fall outside the predicted misspecification interval, not all invariant causes were observed, and we cannot apply the M bound. Otherwise, we may use the M bound.\looseness=-1

\section{Empirical Studies}\label{sec:experiments}

We examine the M-bound and James-bound estimators using synthetic and tobacco consumption data.
With synthetic data, we demonstrate that the M-bound and James-bound estimators produce better estimates in case of misspecification, and show that their misspecification intervals contain the counterfactual outcomes correctly.
Using the tobacco consumption case study, we demonstrate how to collect external data and how to choose between M-bound and James-bound estimators.
We find that the post-intervention California outcomes escape the misspecification error bound, suggesting that there is an actual causal effect.\looseness=-1

\parhead{Implementation Details.} To implement the algorithms we need to manipulate probability distributions and calculate Wasserstein distances with their gradients.
Our implementation expects the input $p_j$ to be non-parametric distributions represented by a collection of atoms and associated probabilities: $p_j = \sum_{x\in X} \delta_{x}\cdot p_j(x)$, where $X$ is the set of atoms and $\delta_x$ is a point mass at $x$. If $p_j$ is discrete, such as a histogram, then the atoms are the possible values of the causes, and $p_j(x)$ their associated probabilities.  If $p_j$ is continuous, then the atoms are samples of $p_j$, and $p_j(x)$ is the normalized density at $x$.
For all experiments, we compute the gradients of
$ (w_1, ..., w_J) \mapsto W_1\bigl(p_0, \sum_j w_j p_j \bigr),$
using the Python Optimal Transport library \citep{flamary2021pot} coupled with PyTorch \citep{pytorch2019}. We use gradient descent with a learning rate $\alpha=5\cdot 10^{-6}$ and $200,000$ epochs.\looseness=-1

\subsection{Experiments with Synthetic Data}
\parhead{Data Description.} We generate synthetic data by defining the conditional distribution $\E[t]{Y | x} = f(x,t)$ and the causes distributions $p_j(x)$. 
We create six different units (called g20, g45, g50, g60, g65, g70), and consider that a single 
cause $x \in \R$ impacts the outcome $Y$. 
The units can be thought of as different groups of people (e.g. cities), and the cause $x$ as the age of each individual in these groups. The six units have different distributions of age (group gX has an average age of X). 
The target group is g45, the panel duration is $T=50$, and the intervention time is $T_0=15$.

\begin{figure}
    \centering
    \includegraphics[width=\linewidth]{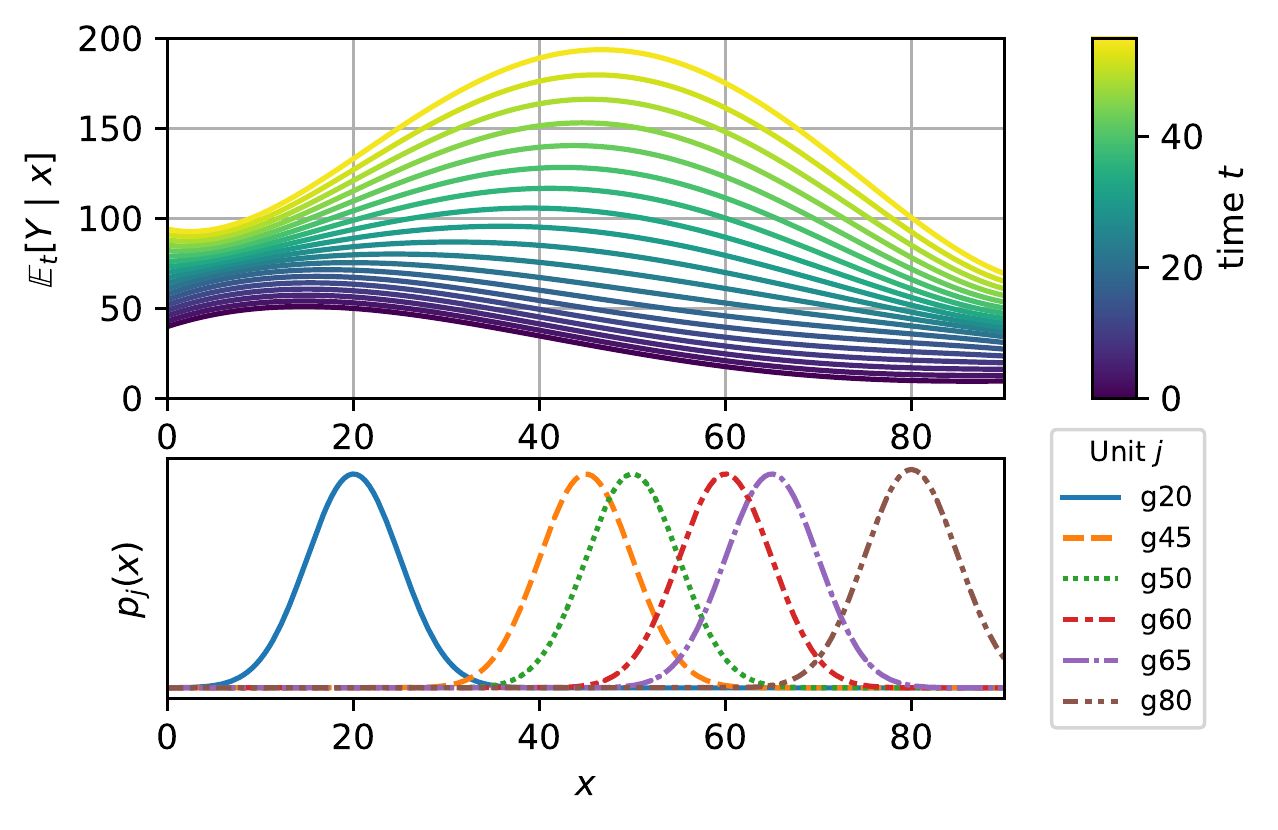}
    \caption{Visualization of the synthetic data generating process. (top) Each colored line represents the expected conditional outcome  $\E[t]{Y | x}$ for a different time $t$ (one line per time), as a function of the cause $x$ (on the x-axis). As time progresses (from darker to lighter), the expected conditional outcomes increase for all values of $x$. For different $x$, the rate of increase over time is different. \mbox{(bottom)} Distributions densities of the causes $x \mapsto p_j(x)$ for each unit $j$. Different lines represent different units. The target unit is g45, which overlaps mostly with unit g50.
    }\label{fig:synthetic_function}
\end{figure}

The closed form equations of $(t,x) \mapsto \E[t]{Y | x}$ and $(j,x) \mapsto p_j(x)$ are in \cref{appsec:experiment} while \Cref{fig:synthetic_function} shows the evolution of $x \mapsto \E[t]{Y | x}$ over time $t$ as well as the distributions $x \mapsto p_j(x)$ for each unit $j$. The expected outcome $\E[t]{Y | x}$ varies over time, in different ways for each $x$. 

\begin{figure}
    \centering
    \includegraphics[width=\linewidth]{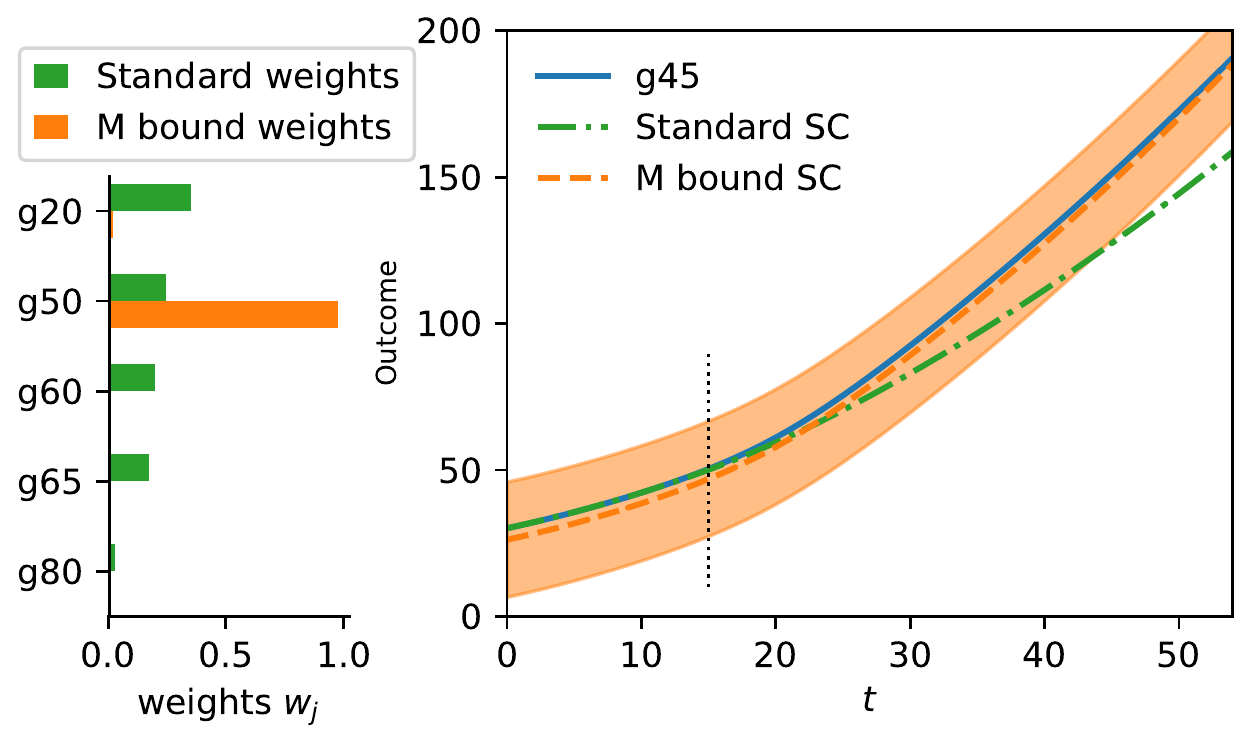}
    \caption{Comparison of the M-bound estimator and the standard SC estimator on synthetic data. (left) Weights returned by each estimator. The M-bound estimator selects donors (g20 and g50) that are most similar to the target (g45). (right) Synthetic outcomes of each estimator, compared to the true outcome. Under misspecification, the M-bound estimator provides more accurate estimates than the standard SC, despite a poorer pre-intervention fit.
}
    \label{fig:synthetic_fit}
\end{figure}

We input the distributions $p_j$ to \cref{alg:mbound} and obtain the weights that minimize the M bound.
As a comparison, we calculate the weights obtained from the standard SC in \cref{eq:estimator}.
We report the weights and the induced synthetic outcomes in \cref{fig:synthetic_fit}. 
Furthermore, we compute $\ell = 4.0$ from $x \mapsto \E[t]{Y\mid x}$ (valid for all $t$). This way, we obtain the exact value of the M-bound and we can form the misspecification interval of \cref{eq:m-interval}, shaded on \cref{fig:synthetic_fit}.

\parhead{Analysis.} 
As shown in \cref{fig:synthetic_fit}, the standard SC places a large weight on donor g20, which is a unit whose individuals are very different from g45 but with similar pre-treatment outcomes. When time increases, the individuals in g20 and g45 evolve differently and the synthetic outcome of the standard SC weights deviates away from the true outcome. 
In contrast, the M-bound estimator places most of the weight mass on the donor g50, which contains individuals with similar $x$  as the target g45. By doing so, the synthetic outcomes might not exactly match the g45 outcomes, but they generalize better over time. 
We also verify that the true outcome is always contained in the misspecification interval (\cref{eq:m-interval}).
We repeat the analysis with the James bound and obtain the same conclusions, reported in \cref{appsec:experiment}.

With external data, we estimated the misspecification error and limited it using the M-bound and James-bound estimators. Without external data, standard SC was incorrect.

\subsection{A Case Study on Real Data}\label{subsec:case-study}
We revisit the tobacco study from \citet{abadie2010synthetic} to illustrate how to collect external data, apply different estimators, and calculate the misspecification intervals.

\parhead{Prop 99. }
A tobacco control program was passed in California in 1988, which increased tobacco taxes by 25 cents.
The tax revenue was used to fund anti-tobacco campaigns.
Our goal is to estimate the causal effect of the tobacco control program on California's tobacco consumption.

The tobacco panel dataset (\cref{fig:data}) is from the \citet{taxburden2019}, which provides the per capita tobacco consumption for 50 states from 1970 to 2019.
The intervention of interest is the tobacco program, Prop 99.
The observed outcomes for California after 1988 are under intervention.
All the other observed outcomes in the dataset are assumed to be under no intervention.

\parhead{External Data Collection.}
First, we identify the potential causes of smoking.
According to \citet{turner2004individual}, smoking is heavily influenced by societal and cultural factors.
While these factors are difficult to measure directly, they are often correlated with demographics. 
Several studies have found that cigarette consumption varies significantly by age, gender, race, and ethnicity \citep{sakuma2016tobacco, cornelius2022tobacco}. 
As a result, we use \emph{age}, \emph{sex}, and \emph{ethnicity/race} as proxies for the causes of smoking.

We use the American Community Survey (ACS) to formulate a distribution of causes for each unit. The ACS is a demographics survey program conducted continuously by the U.S. Census Bureau \citep{censusbureau2020acs}. 
It reports population demographics at different geographical scales, from city boroughs to states. We accessed the ACS data with the Census Reporter API \citep{censusreporter}. 
For each state, the ACS provides the joint distribution of the variables \emph{age, race, sex}. Each variable is discretized into multiple bins: \textit{age} into 14 bins (e.g. 15 to 17, 20 to 24 years old), \textit{race} takes 8 values (Asian, Black, Native American, Pacific Islander, White non-Hispanic, White Hispanic, Mix, and Other), and \textit{sex} takes 2 values (Male, Female).
The joints $x \mapsto p_j(x)$ over these variables are defined for each state on these $14 \times 8\times2=224$ demographics combinations (atoms).\looseness=-1

We estimate $\ell$ using additional survey data from the Tobacco Use Supplement to the Current Population Survey. This independent study collects individual demographic information along with tobacco consumption. We form the expected tobacco consumption given each invariant cause and compute the induced $\ell$. More details about the computation of Lipschitz constant can be found in \cref{appsec:experiment}.


\parhead{The M-bound Estimator.} The M-bound estimator uses our newly formed distributions $p_0, ..., p_J$ to compute a set of SC weights. We report the weights in \cref{app:weights} and the SC outcomes with the misspecification interval in \cref{fig:california_jamesbound}. Among the set of 50 potential donors, five obtained non-zero weights: New Mexico, Nevada, D.C, Hawaii and Texas.  As expected, the M-bound estimator selected states that are similar to California. New Mexico and Nevada are geographically close and have similar demographics. Both D.C. and California have a relatively young active population. And, California is the number one destination for Hawaiians moving to the US mainland (from US census).

In \cref{fig:california_jamesbound}, the solid and dotted lines denote the observed and synthetic California outcomes.
The shaded areas represent the misspecification intervals. California pre-intervention outcomes fall within the estimated M-bound interval, but synthetic California is not a perfect fit, there is misspecification. In spite of the misspecification, \cref{fig:california_jamesbound} shows that the post-intervention outcomes are outside of the bounds, suggesting a causal effect.

\parhead{The James-bound Estimator.} As discussed in \cref{sec:choosing}, the M-bound misspecification interval is only valid if we observe all the invariant causes. According to \cref{fig:california_jamesbound},  California's pre-intervention outcomes fall within the M-bound intervals. 
We find, however, that when some other states are considered as the target unit, their observed outcomes before the intervention are not always within the interval.  

\begin{figure}[t]
    \centering
    \includegraphics[width=\linewidth]{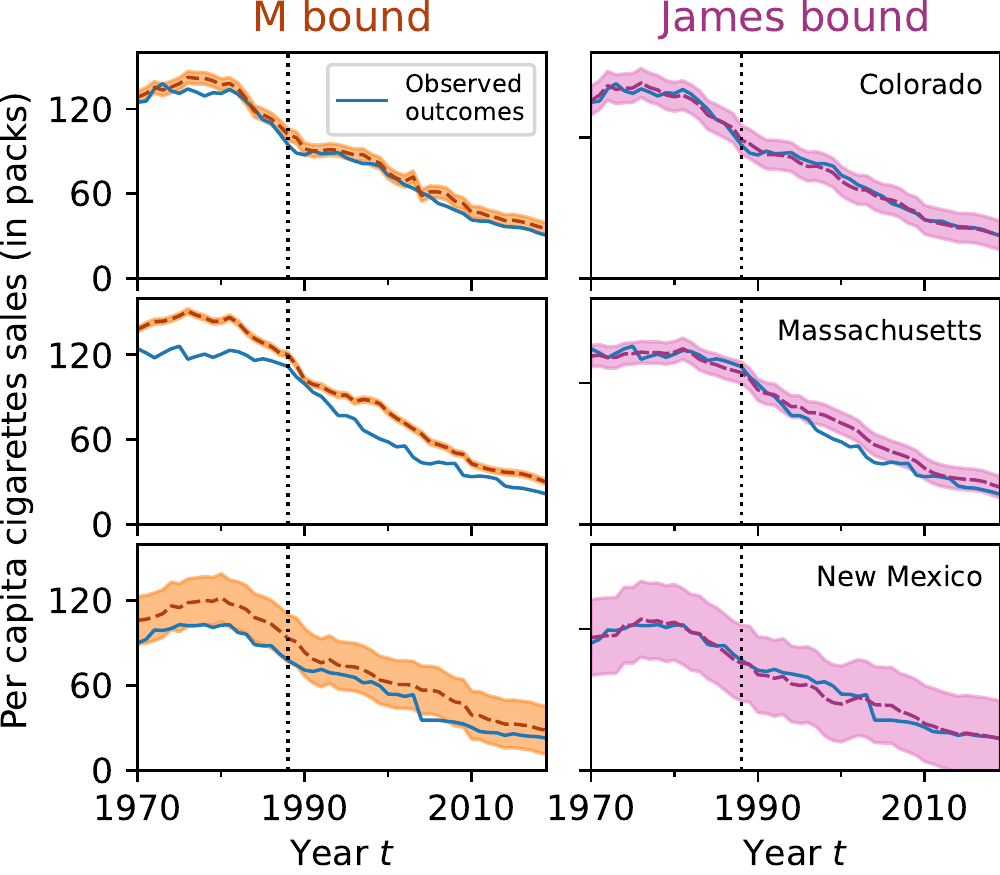}
    \caption{Placebo study of the M-bound estimator (left) and the James-bound estimator (right), on Colorado, Massachusetts, and New Mexico. The M-bound synthetic outcomes are outside of the
misspecification interval before the intervention. This suggests that not all invariant causes are observed and that the James bound should be used. The James-bound estimator accounts for the missing causes, with wider misspecification intervals.}
    \label{fig:james-bounds}
\end{figure}

We perform \textit{placebo tests} \citep{abadie2010synthetic} where each donor is considered to be the target and a synthetic control is constructed using the other donors. 
Because the donors did not receive the intervention, we expect synthetic outcomes to match observed outcomes.
In \cref{fig:james-bounds}, we illustrate the comparisons for three states,  Colorado, Massachusetts, New Mexico.
For comparisons on all states, see \cref{appsec:experiment}.

\cref{fig:james-bounds} (left) shows the synthetic outcome estimates by the M-bound estimator.
Both Colorado and Massachusetts's pre-intervention outcomes are outside of the misspecification interval. This suggests that not all invariant causes are observed.
While New Mexico's pre-intervention outcomes lie within the misspecification interval of the synthetic New Mexico, the error bound is too large to use synthetic control.

\cref{fig:james-bounds} (right) shows the synthetic outcome estimates using the James-bound estimator. 
We observe that the pre-intervention outcomes across states now all fall within the James-bound misspecification intervals, which are also wider than the M-bound intervals.
After the intervention, the observed tobacco consumption in Colorado remains in the James-bound misspecification interval, suggesting the intervention had no effect. This is expected because Colorado did not implement an anti-tobacco program like California.
For Massachusetts, the James-bound interval is narrow enough to detect a decrease in tobacco consumption that is not due to misspecification. In fact, this is consistent with the policies taken by this state in 1993 to raise taxes and increase its Massachusetts Tobacco Control Program.

The placebo test provides further evidence that the tobacco control program in California indeed had a causal effect on tobacco consumption.
In states without tobacco control programs, their outcomes fall within the misspecification interval, whereas California's outcome does not.

\section{Discussion}

We address the problem of the misspecification of linear assumptions in synthetic controls. We relax assumptions commonly assumed in the literature (A1, A2), derive two misspecification bounds, and propose corresponding estimators.
The key idea is to leverage external data to bound and minimize misspecification.
Each bound comes with requirements: data must be available for the M bound (observe all causes), and we identify a modeling assumption for the James bound (A3).
As an area of future research, we can explore other SC estimation procedures that might be enabled by these two requirements.

\clearpage

\onecolumn
\printbibliography

\newpage
\appendix
\onecolumn
\makeatletter
\newtheorem{repeatbound@}{Bound}
\newenvironment{repeatbound}[1]{%
    \def\therepeatbound@{\ref{#1}}
    \repeatbound@
}
{\endrepeatbound@}

\makeatother

\begin{appendices}

\section{Technical Details}\label{appsec:technical}

\subsection{Lipschitz function}
A function $f: \R^n \rightarrow \R$ is said to be $\ell$-Lipschitz if 
\begin{equation}
\forall (x,y) \in (\R^n)^2, \quad |f(x) - f(y) | \leq \ell \cdot \| x - y\|_1
\end{equation}
More precisely, the function $f$ can be from and to any metric spaces, with their associated distance. 
Here we use the $L_1$ norm to measure the distance between $x$ and $y$ in $\R^n$, and we use the absolute value to measure the distance between $f(x)$ and $f(y)$ in $\R$. 

A Lipschitz function is limited in how fast it can change. In the context of synthetic control, if $\ell$ is small, it implies that a change in the causes induces a small change in the outcomes. Having a small $\ell$ suggests that misspecification of the causes will have a limited impact on the outcome.

\subsection{Proof of the M Bound}
\begin{repeatbound}{theorem:bound1}[M bound]
For any $t$, let assume that $x \mapsto \E[t]{Y | x}$ is $\ell$-Lipschitz, then for any weights in the simplex $w$, we have the Misspecification error bound (M-bound):
\begin{align}
\abs{
    \E{Y_{0t}} - \sum_{j=1}^J w_j \E{Y_{jt}}
} \leq  \ell \cdot W_1\parens{p_{0}, \hat p_0},
\end{align}
where $\hat p_0 = \sum_j w_j p_{j}$ and $W_1$ is a $\ell_1$-Wasserstein \mbox{distance}.
\end{repeatbound}
\begin{proof}
Notice that for any unit $j \in \bbrackets{0,J}$, the expected outcome writes $\E{Y_{jt}} = \int_x \E[t]{Y | X=x}p_j(x)\text{d}x $.

Fix some weights $w_{1:J} \in \Delta^J$ and then by the linearity of the integral,
\begin{align*}
\E{Y_{0t}} - \sum_{j=1}^J w_j \E{Y_{jt}} 
& = \int_x \E[t]{Y | x}\textstyle \left(p_0(x) - \sum_j w_j p_j(x) \right)\text{d}x \\
&= \int_x \E[t]{Y | x}\textstyle \left(p_0(x) - \hat p_0(x) \right)\text{d}x \\
& \leq W_1(p_0, \hat p_0) \cdot \ell
\end{align*}
where the inequality comes from Kantorovich duality (Theorem 5.10, \citet{villani2009optimal}) about Wasserstein distances.
\end{proof}

\subsection{Proof of the James Bound}

\begin{repeatbound}{theorem:bound2}[James bound]
For $t \geq T_0$, let assume that $x \mapsto \E[t]{Y | x}$ is $\ell$-Lipschitz, then for any weights in the simplex $w \in \Delta^J$, we have Just Another Misspecification Errors bound (James bound) :

\begin{dmath*}
\abs{
    \E{Y_{0t}} - \sum_{j=1}^J w_j \E{Y_{jt}}
} \leq 
 \ell \cdot W_1(p_0(x), \hat p_0(x)) +\max_{u < T_0} \abs{\E{Y_{0u}} - \sum_{j=1}^J w_j \E{Y_{ju}}}  + {\inf_{\alpha \in \Delta^{ T_0}} \abs{\int_{z} \textstyle  \Bigl(p_0(z) - \hat p_0(z) \Bigr) \Bigl(\E[t]{Y | z} - \sum\limits_{u<T_0} \alpha_u \E[u]{Y | z} \Bigr) \textnormal{d}z}}. 
\end{dmath*}
\end{repeatbound}
Before proving the James bound, we need one lemma which follows from assumption A3.

\begin{lemma}
Suppose the distributions of causes $(x,z) \mapsto p_j(x,z)$ and the expected outcomes $(x,z) \mapsto \E[t]{Y|x,z}$ satisfy assumption A3, then $\E[t]{Y|x} = g_t(x) + constant$ and $\E[t]{Y|z} = h_t(z) + constant$ where the $constant$ terms are independent of $x$ and $z$ (they can depend on $t$).
\end{lemma}

\begin{proof} Fix $t$ and $j$, and define the function $f$ by  $f_t(x,z) = \E[t]{Y \mid x,z}$.
We suppose A3, that is $p_j(x,z) = p_j(x)p_j(z)$ and that there exist two functions $g_t$ and $h_t$ such that
\[ 
\forall (x,z), ~ f_t(x,z) = g_t(x) + h_t(z).
\]

Then
\begin{align*}
    \E[t]{Y|x} &= \E[p_j(z \mid x)]{\E[t]{Y\mid x,z} \mid x} \\
    &= \int_z \parens{g_t(x) + h_t(z)} p_j(z) \text{d}z \\
    &= g_t(x) + \int_z h_t(z) p_j(z)\text{d}z.
\end{align*}
We notice that $\int_z h_t(z) p_j(z)\text{d}z$ is a constant independent of $x$ or $z$. (It is the expectation of $z \mapsto h_t(z)$ with respect to $z\mapsto p_j(z)$.)

We obtain a similar result for $ \E[t]{Y|z}$, $h_t(z)$ and the constant $\int_z g_t(x) p_j(x)\text{d}x$.
\end{proof}

We can now prove the James bound.
\begin{proof}
Fix some weights $w \in \Delta^J$, weights $\alpha \in \Delta^{T_0}$ then,

\begin{align*}
\E{Y_{0t}} - \sum_{j=1}^J w_j \E{Y_{jt}} 
& = \int_{(x,z)} \E[t]{Y | x, z}\textstyle \left(p_0(x,z) - \sum_j w_j p_j(x,z) \right)\text{d}x\text{d}z \\
& = \int_{(x,z)} (g_t(x)+h_t(z))\textstyle \left(p_0(x)p_0(z) - \sum_j w_j p_j(x)p_j(z) \right)\text{d}x\text{d}z \\
&= \underbrace{\int_x g_t(x)\textstyle \left(p_0(x) -  \sum_j w_j p_j(x)\right)\text{d}x}_{A} + \underbrace{\int_z h_t(z)\textstyle \left(p_0(z) -  \sum_j w_j p_j(z) \right)\text{d}z}_{B}.
\end{align*}

We have $g_t(x) = \E[t]{Y|x} + {constant}$, the constant cancels out in $A$ and we obtain:
$$ |A| = \abs{\int_x \E[t]{Y|x} \textstyle \left(p_0(x) -  \sum_j w_j p_j(x)\right)\text{d}x} \leq \ell \cdot W_1(p_0(x), \hat p_0(x) ). $$ 

We have $h_t(z) = \E[t]{Y|z} + {constant}$, the constant cancels in $B$ and we obtain:
\begin{align*}
B &= \int_z \E[t]{Y|z} \textstyle \left(p_0(z) -  \hat p_0(z) \right)\text{d}z
\end{align*}
For any $\alpha \in \Delta^{T_0}$ we have:

\begin{align*}
B &= \int_z \Bigl({\textstyle\sum\limits_{~u<T_0}} \alpha_u \E[u]{Y | z} \Bigr)  \left(p_0(z) -  \hat p_0(z) \right)\text{d}z +  \int_z \Bigl(\E[t]{Y | z} - {\textstyle\sum\limits_{u<T_0}} \alpha_u \E[u]{Y | z} \Bigr) \left(p_0(z) -  \hat p_0(z) \right)\text{d}z\\
&=\sum\limits_{u<T_0} \alpha_u \int_z \Bigl(\E[u]{Y | z}p_0(z) - {\textstyle\sum_j} w_j \E[u]{Y |z}p_j(z)\Bigr) \text{d}z + \int_z \Bigl(\E[t]{Y | z} - {\textstyle\sum\limits_{u<T_0}} \alpha_u \E[u]{Y | z} \Bigr) \left(p_0(z) -  \hat p_0(z) \right)\text{d}z \\
&=\sum\limits_{u<T_0} \alpha_u \Bigl( \E{Y_{0u}} - {\textstyle\sum_j} w_j \E{Y_{ju}} \Bigr) \text{d}z + \int_z \Bigl(\E[t]{Y | z} - {\textstyle\sum\limits_{u<T_0}} \alpha_u \E[u]{Y | z} \Bigr) \left(p_0(z) -  \hat p_0(z) \right)\text{d}z.
\end{align*}

So, 
\begin{align*}
    |B| &\leq \sum\limits_{u<T_0} \alpha_u \Bigl| \E{Y_{0u}} - {\textstyle\sum_j} w_j \E{Y_{ju}} \Bigr| \text{d}z + \left|\int_z \Bigl(\E[t]{Y | z} - {\textstyle\sum\limits_{u<T_0}} \alpha_u \E[u]{Y | z} \Bigr) \left(p_0(z) -  \hat p_0(z) \right)\text{d}z\right|\\
    &\leq \parens{\sum\limits_{u<T_0} \alpha_u} \max\limits_{u<T_0} \Bigl| \E{Y_{0u}} - {\textstyle\sum_j} w_j \E{Y_{ju}} \Bigr| \text{d}z + \left|\int_z \Bigl(\E[t]{Y | z} - {\textstyle\sum\limits_{u<T_0}} \alpha_u \E[u]{Y | z} \Bigr) \left(p_0(z) -  \hat p_0(z) \right)\text{d}z\right|\\
    &=\max\limits_{u<T_0} \Bigl| \E{Y_{0u}} - {\textstyle\sum_j} w_j \E{Y_{ju}} \Bigr| \text{d}z + \left|\int_z \Bigl(\E[t]{Y | z} - {\textstyle\sum\limits_{u<T_0}} \alpha_u \E[u]{Y | z} \Bigr) \left(p_0(z) -  \hat p_0(z) \right)\text{d}z\right|.
\end{align*}

Because the previous inequality hold for any $\alpha \in \Delta^{T_0}$, we can ``take'' the inf on the right term.

We obtain, 
\begin{align*}
    |B| &\leq \max\limits_{u<T_0} \Bigl| \E{Y_{0u}} - {\textstyle\sum_j} w_j \E{Y_{ju}} \Bigr| \text{d}z + \inf_{\alpha \in \Delta^{T_0}} \left|\int_z \Bigl(\E[t]{Y | z} -  {\textstyle\sum\limits_{u<T_0}} \alpha_u \E[u]{Y | z} \Bigr) \left(p_0(z) -  \hat p_0(z) \right)\text{d}z\right|.
\end{align*}

This proves the James bound.
\end{proof}

\subsection{Interpretation of \cref{eqn:bound2-3}}

In the James bound, we can compute the following terms:
\begin{itemize}
    \item $\ell \cdot W_1(p_0(x), \hat p_0(x))$ is estimated with the external data on the subset of observed causes.
    \item $\max\limits_{u<T_0} \Bigl| \E{Y_{0u}} - {\textstyle\sum_j} w_j \E{Y_{ju}} \Bigr| \text{d}z$ is estimated from the outcome data.
\end{itemize}
The last term: $\inf\limits_{\alpha \in \Delta^{T_0}} \left|{\displaystyle\int_z} \Bigl(p_0(z) -  \sum\limits_{j} w_j p_j(z) \Bigr) \Bigl(\E[t]{Y | z} -  {\textstyle\sum\limits_{u<T_0}} \alpha_u \E[u]{Y | z} \Bigr) \text{d}z\right|$ cannot be estimated from observed data. Hence, we defined the James-bound estimator without this term. and showed that it was minimizing the James bound only if this last term is negligible.

We give justification as to why this last term might be negligible, at least in comparison to the two other terms.
If any of the two following conditions holds, the last term is 0:
\begin{enumerate}
    \item If $p_0 = \sum_j w_j p_j $.
    \item If there exists $\alpha \in \Delta^{T_0}$ such that $\E[t]{Y | z} -  {\textstyle\sum\limits_{u<T_0}} \alpha_u \E[u]{Y | z} = constant$.
\end{enumerate}

Naturally, we are not expecting the first condition to hold, but at least we can hope that $p_0(z) - \sum_j w_j p_j(z)$ is of the same order of magnitude as $p_0(x) - \sum_j w_j p_j(x)$ (and so of $W_1(p_0(x), \hat p_0(x))$, the first term). If in addition $\E[t]{Y | z} -  {\textstyle\sum\limits_{u<T_0}} \alpha_u \E[u]{Y | z}$ is small, then the last term (which is a product of two small terms) is negligible compared to the other terms of the James bound.

With this intuition, it seems possible to finding a $\alpha$ that makes the full integral close to zero should be possible. 

More concretely, we give two examples of models for which the term (\ref{eqn:bound2-3}) is null.

\paragraph{Standard SC setting}
In the standard SC setting, assumption A2 is made. The practitioner assumes that there exists a set of weights $(w_j)$ such that $p_0 = \sum_j w_j p_j$ (with the notations of factor models, each $p_j$ is a point mass located at the latent factor $\mu_j$, such that $\mu_0 = \sum_j w_j \mu_j$).
In particular, it implies that $\Bigl(p_0(z) -  \sum\limits_{j} w_j p_j(z) \Bigr) = 0$ and so $(\ref{eqn:bound2-3}) = 0$.

\paragraph{Arbitrary model with linear conditional expectation.}
We now assume arbitrary distributions $p_j$. They can be point mass on linear factors as in standard SC or arbitrary continuous distributions. A2 does not need to hold, it may be impossible to write the target as a linear combination of the donors. However, we assume that the response functions $(z \mapsto \E[t]{Y\mid z})_t$ are of the form $\E[t]{Y\mid z} = \beta_t^\top z$, with $\beta_t$ being able to change arbitrarily over time. Actually, we even requires the $\beta_t$ to change enough such that there exists $\alpha \in \Delta^{T_0}$ such that they are linearly independent and there exists $\beta_t =  \sum\limits_{u<T_0} \alpha_u \beta_u$. In that case again, $(\ref{eqn:bound2-3}) = 0$.

\subsection{The James-bound Estimator}
We adapt the M-bound estimator algorithm from \cref{alg:mbound} into \cref{alg:james}.
\begin{algorithm}
   \caption{Minimization of the James-bound}
   \label{alg:james}
\begin{algorithmic}
   \STATE {\bfseries Input:} Distributions $p_0, ..., p_J$; Pre-intervention measurements $\{(y_{0t}, ..., y_{jt})_{t=0, ...,T_0-1}\}$;
   learning rate $\alpha$; number of epochs $E$, James parameter $\lambda$.
   \STATE {\bfseries Output:} $(w_j)$ minimizing the James-bound. 
   \STATE $(w_1, ..., w_J) \leftarrow \bigl(\frac1J, ..., \frac1J\bigr) $
   \FOR{$e=1$ {\bfseries to} $E$}
       \STATE $\hat p_0 \leftarrow \sum w_j p_j$
       \STATE $\text{grad} \leftarrow \nabla_{w} \Bigl(\max\limits_{t < T_0} \bigl|y_{0t} - \sum\limits_{j=1}^J w_jm_{jt}\bigr| +  \lambda \cdot W_1\bigl(p_0, \hat p_0 \bigr)  \Bigr) $
       \STATE $w \leftarrow w - \alpha \cdot \text{grad}$
       \STATE $w \leftarrow \text{project\_simplex}(w) $
   \ENDFOR
  \STATE \textbf{return} $w$
\end{algorithmic}
\end{algorithm}

\section{Experiment Details}\label{appsec:experiment}


\subsection{Simulation Details}
For the synthetic experiment, we generate the outcomes under no intervention by defining the conditional expected outcomes $f: (x,t) \mapsto \E[t]{Y|x}$ and the unit specific causes distributions $x \mapsto p_j(x)$. In this experiment, $x$ is a single scalar variable.

The function $f$ we choose is represented in \cref{fig:synthetic_function} (top). It enjoys a closed-form expression:
\begin{dmath*}f(x,t) = - \frac{13 t^{2} x^{4}}{2100000000} + \frac{71 t^{2} x^{3}}{78750000} - \frac{10141 t^{2} x^{2}}{252000000} + \frac{12521 t^{2} x}{12600000} + t + 4.07142857142857 \cdot 10^{-6} x^{4} \log{\left(e^{\frac{t}{3} - \frac{20}{3}} + 1 \right)} - \frac{43 x^{4}}{13300000} - 0.000731428571428571 x^{3} \log{\left(e^{\frac{t}{3} - \frac{20}{3}} + 1 \right)} + \frac{1313 x^{3}}{1496250} + 0.0380053571428571 x^{2} \log{\left(e^{\frac{t}{3} - \frac{20}{3}} + 1 \right)} - \frac{359953 x^{2}}{4788000} - 0.500107142857143 x \log{\left(e^{\frac{t}{3} - \frac{20}{3}} + 1 \right)} + \frac{401813 x}{239400} + 40\end{dmath*}.

It was generated by combining Lagrange polynomials in $x$ with time varying coefficients.

For each group gXX (g20, g45, g50, g60, g65, g70), their associated distribution of causes is given by a normal distribution centered at XX (e.g. at 20 for g20), and with scale 5 (variance 25). Each distribution is represented in \cref{fig:synthetic_function} (bottom).

Because our implementations of the M-bound estimator and James-bound estimator use non-parametric distributions represented by a collection of atoms and associated probabilities, each $p_j$ is more precisely defined as  $$p_j \propto \sum_{x\in X} \delta_{x}\cdot \mathcal{N}(x; \mu_j, 5^2)$$ where $\mu_j=$ XX for each group gXX, and the set of atoms $X$ is $X = \{90\cdot k/199 \mid k \in \bbrackets{0,199}\}$. 

\subsection{Evaluation of the James-bound Estimator on Synthetic Data}
\begin{figure}[h]
    \centering
    \begin{subfigure}[b]{0.45\textwidth}
         \centering
         \includegraphics[width=\textwidth]{fig/synthetic_fit.pdf}
         \caption{M-bound estimator.}
     \end{subfigure}
     \hfill
     \begin{subfigure}[b]{0.45\textwidth}
         \centering
         \includegraphics[width=\textwidth]{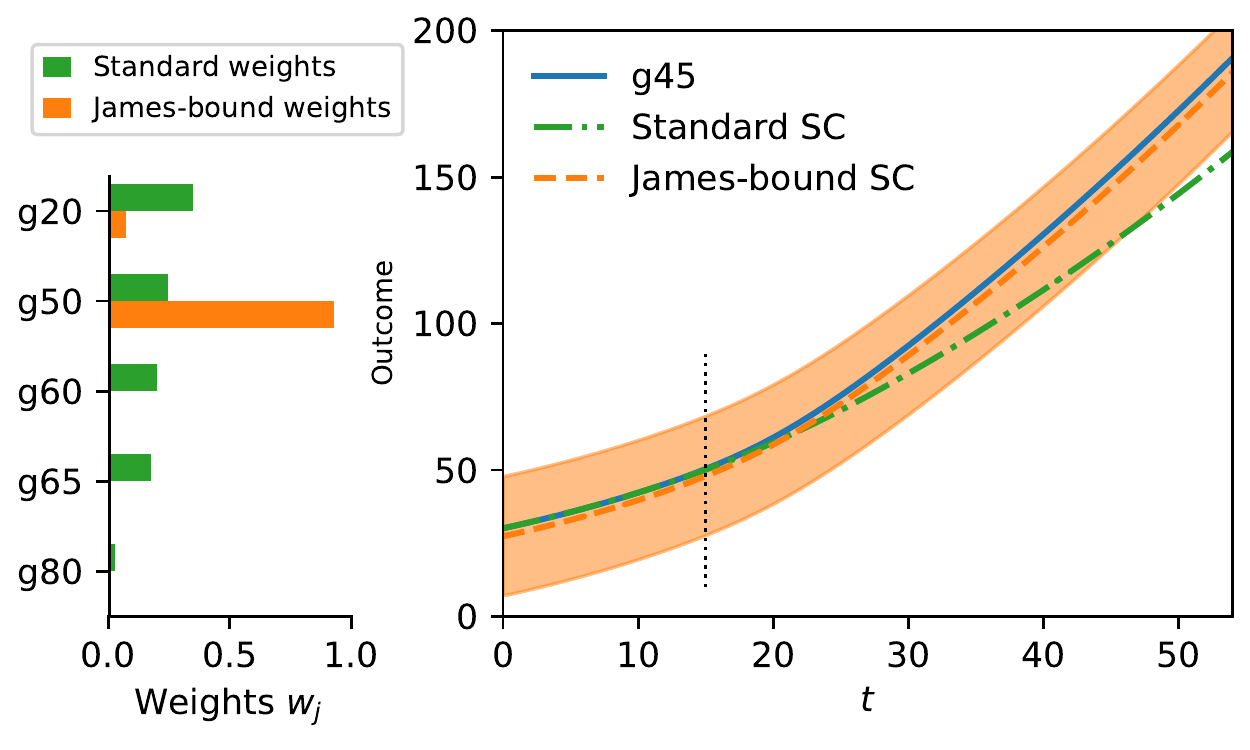}
         \caption{James-bound estimator.}
     \end{subfigure}
    \caption{Comparison of the M-bound estimator, the James-bound estimator and the
standard SC estimator on the synthetic data. Both the James-bound and M-bound estimators produce more accurate counterfactual estimates than the standard SC, despite a poorer pre-intervention fit. The M-bound estimator favors donor g50 (which is the unit with individuals most similar to the target). The standard SC estimator favors donor g20, which has similar outcomes before the intervention but have different individuals. The James-bound estimator trades off and selects mostly g50 with a little of g20. Both the M-bound and James-bound misspecification intervals contain the true outcomes. }
    \label{fig:app-jamesbound}
\end{figure}

\cref{fig:app-jamesbound} reports the estimates and weights produced by the M-bound, James-bound, and standard SC estimators.
Both M-bound and James-bound estimators select donor units that are more similar to the target.
The M-bound estimator favors donor g50, a unit with individuals most similar to the target. 
Using the standard SC estimator, donor g20 is preferred, as it has similar outcomes, but different individuals, before the intervention.
The James-bound estimator chooses mainly donor g50 with a small selection of donor g20 as it trades off between selecting donors with similar outcomes and similar individuals.

Both the M-bound and James-bound estimators produce misspecification intervals that cover the true outcomes. As expected, the James bound estimator produces a wider misspecification interval than the M bound.
The James-bound estimator also produces a better fit for the observed data than the M-bound estimator. This is expected since the M-bound estimator does not consider the pre-intervention outcomes, whereas the James-bound estimator does.

\subsection{Using Survey Data to Estimate the Lipschitz Constant }
To compute $\ell$ for the tobacco case study, we leverage external survey data. The (smallest) Lipschitz constant of a function $f:\R^n \rightarrow \R$ is by definition, $$ \inf_{x \neq x' } \frac{|f(x) - f(x')|}{\|x - x'\|_1}.$$ (We use the $L_1$ norm over $\R^n$).

For the M and James bounds, we need to compute the Lipschitz constant of $x \mapsto \E[t]{Y|x}$, that is, of the expected tobacco consumption given the causes $x$.
We use additional survey data from the Tobacco Use Supplement to the Current Population Survey (TUS-CPS). This study collects individual demographic information along with tobacco consumption. We estimate the expected tobacco consumption given the invariant causes $x$ and compute the induced $\ell$ with the formula above by computing the pairwise differences, normalized by the differences of causes $x$.

\parhead{Handling categorical causes.}
For both the Wasserstein distance and the Lipschitz constant, we take $L_1$ norms over the causes $x$. Some causes might be categorical. We represent a categorical variable $C$ which can take $k$ values $c_1,..., c_k$ as a one-half-hot encoding with $k$ different binary variables $x_1, ..., x_k$ with values $0$ and $\frac12$, such that $C=c_r$ is represented by $(x_1,...x_k) = ( \mathds{1}(i=r))_{1\leq i \leq k}$. 

\newpage
\subsection{California M-estimator Weights}
\label{app:weights}
\begin{table}[h!]
    \centering
    \begin{tabular}{ccccc}
    \toprule
        D.C & Hawaii & Nevada & New Mexico & Texas \\
        \midrule
        0.106 & 0.166 & 0.195 & 0.209 & 0.324\\
    \bottomrule
    \end{tabular}
    \caption{Non-zero weights returned by the M-bound estimator for the  synthetic California of \cref{fig:california_jamesbound}.}
    \label{tab:Mweights}
\end{table}

\subsection{Placebo tests}
In \cref{fig:placebo-part1,fig:placebo-part2,fig:placebo-part3}, we report the full placebo study with all the states.
\renewcommand{\thefigure}{\arabic{figure}a}
\begin{figure}
    \centering
    \includegraphics[width=\linewidth]{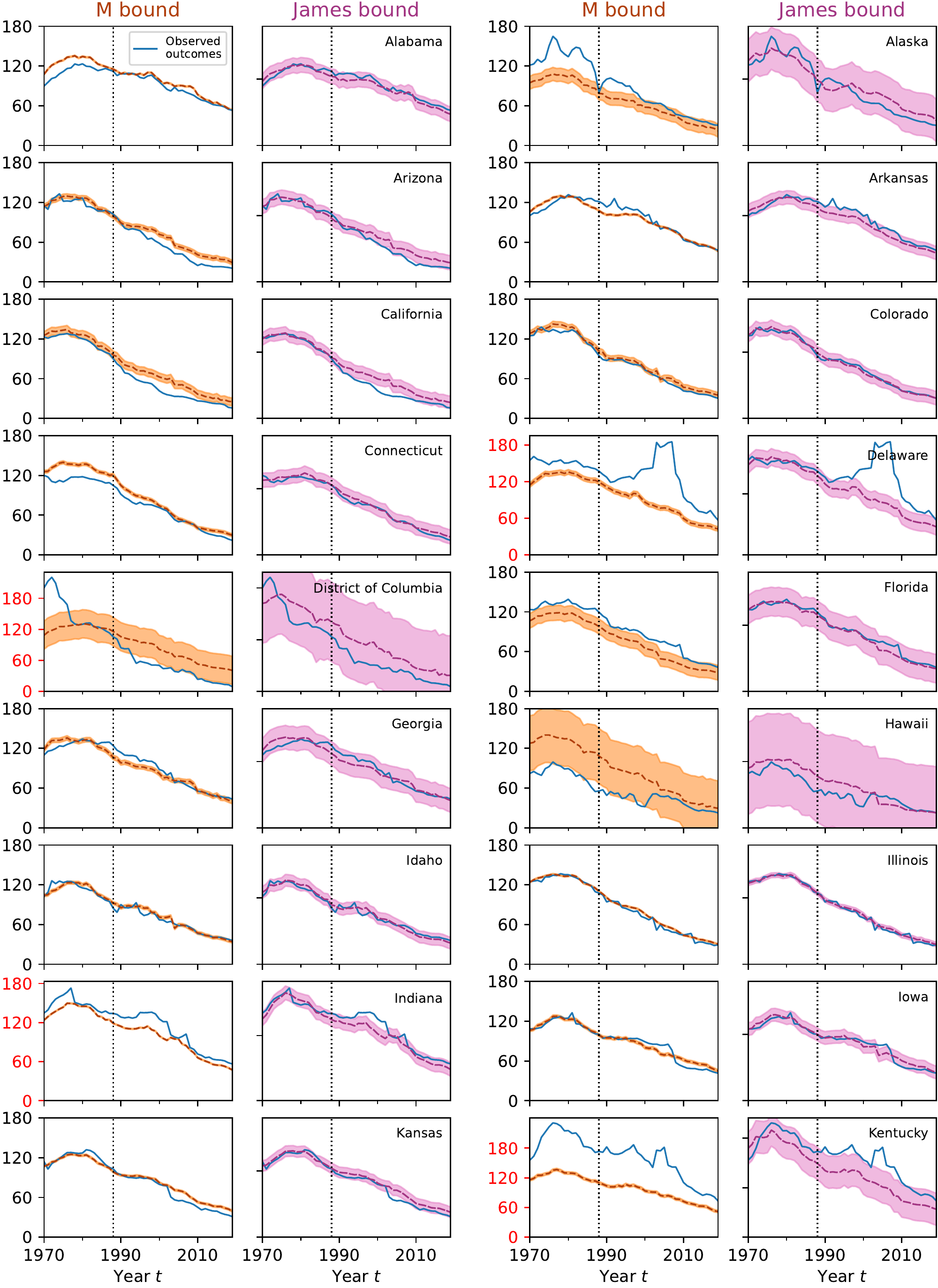}
    \caption{Placebo study, part 1, of the M-bound estimator (left) and the James-bound estimator (right). The y-axis represents the per capita cigarette sales (in packs). The y-axis usually spans from 0 to 180 and is colored in red otherwise.}
    \label{fig:placebo-part1}
\end{figure}
\addtocounter{figure}{-1}
\renewcommand{\thefigure}{\arabic{figure}b}
\begin{figure}
    \centering
    \includegraphics[width=\linewidth]{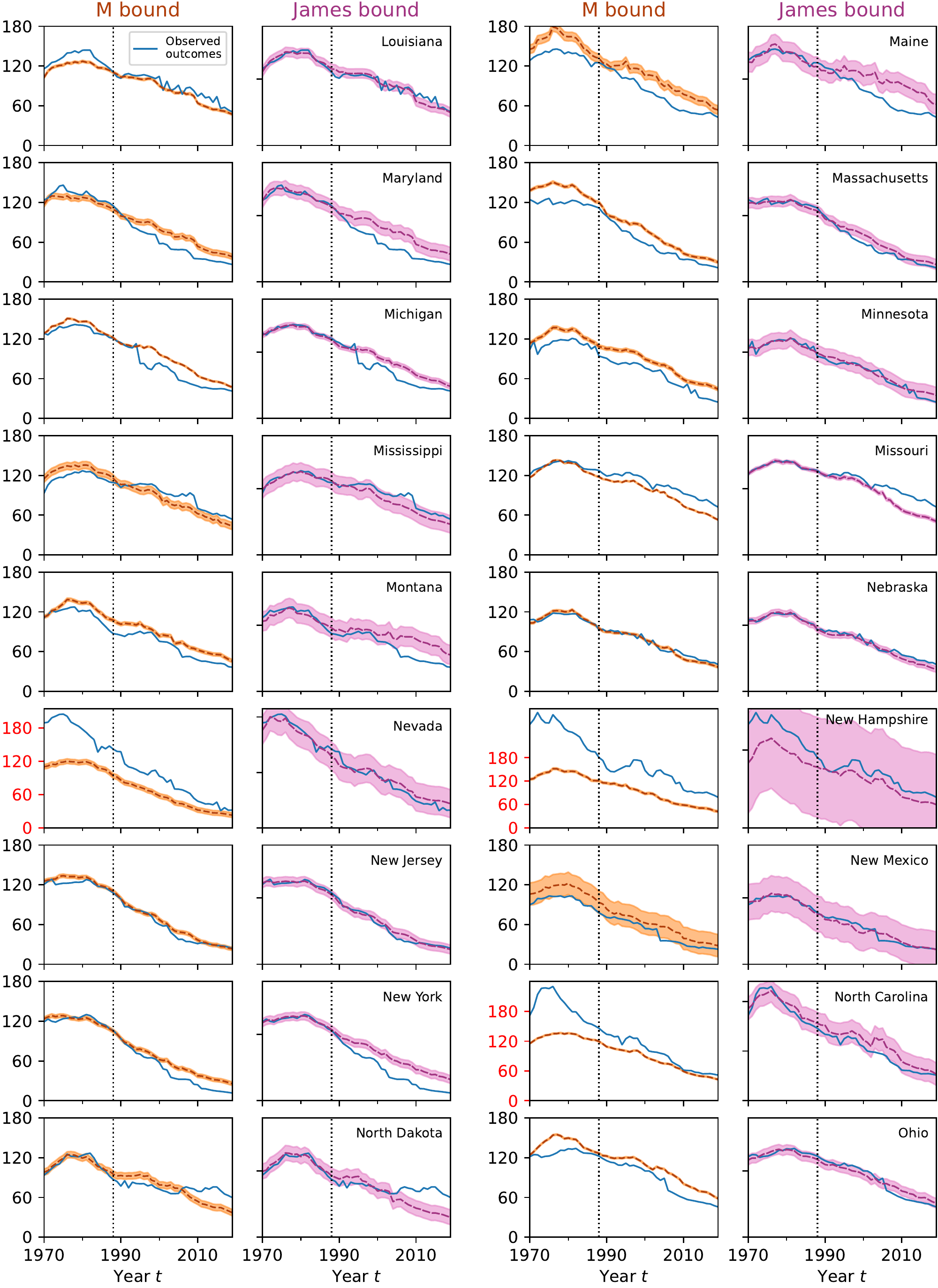}
    \caption{Placebo study, part 2, of the M-bound estimator (left) and the James-bound estimator (right). The y-axis represents the per capita cigarette sales (in packs). The y-axis usually spans from 0 to 180 and is colored in red otherwise.}
    \label{fig:placebo-part2}
\end{figure}
\addtocounter{figure}{-1}
\renewcommand{\thefigure}{\arabic{figure}c}
\begin{figure}
    \centering
    \includegraphics[width=\linewidth]{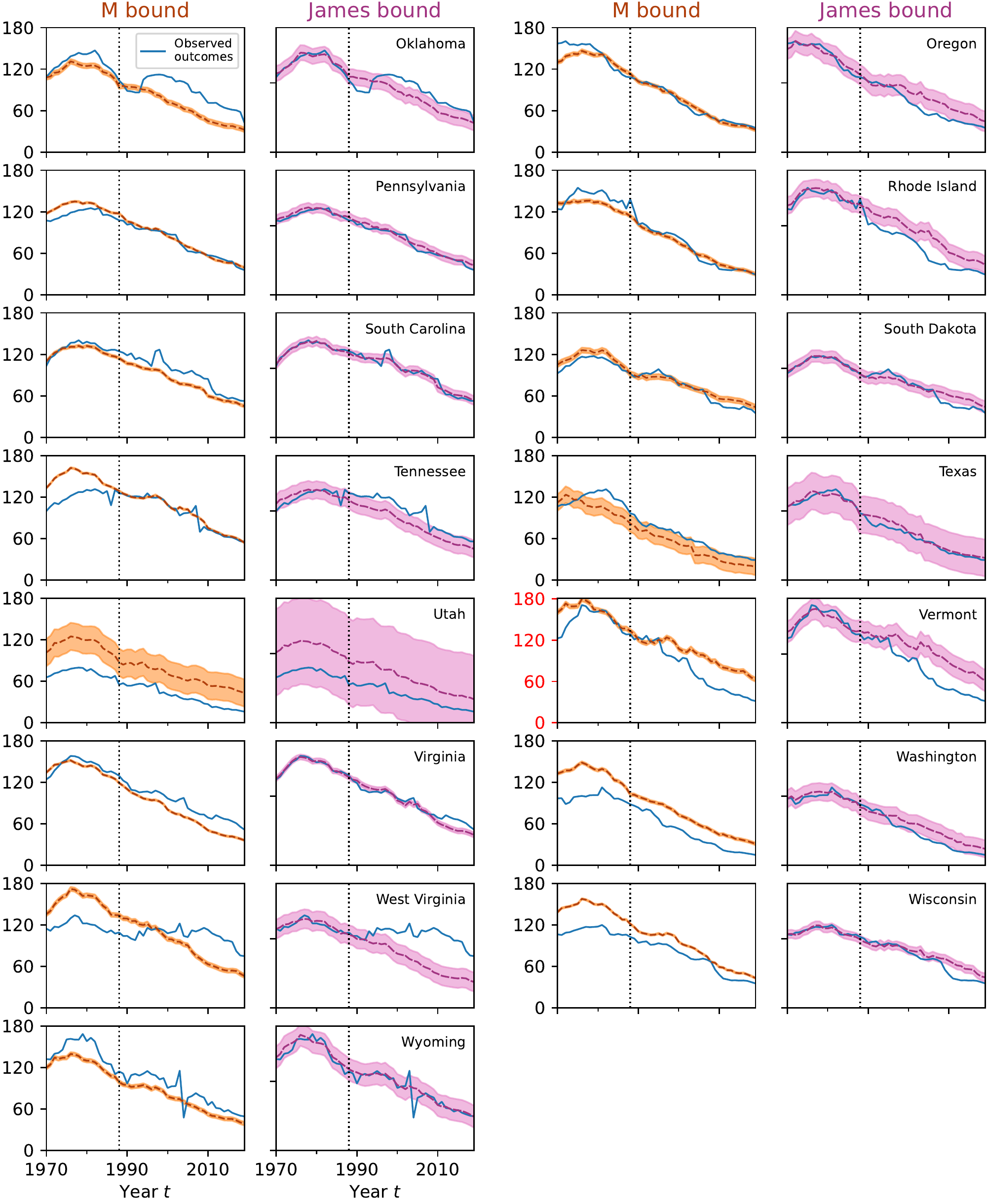}
    \caption{Placebo study, part 3, of the M-bound estimator (left) and the James-bound estimator (right). The y-axis represents the per capita cigarette sales (in packs). The y-axis usually spans from 0 to 180 and is colored in red otherwise.}
    \label{fig:placebo-part3}
\end{figure}
\renewcommand{\thefigure}{\arabic{figure}}

\end{appendices}

\end{document}